\documentclass[runningheads]{svmult}
\input epsf
\usepackage{makeidx}   
\usepackage{graphicx}  
\usepackage{subeqnar}  
\usepackage{cropmark} 
\usepackage{physprbb}  
\begin{document}
\title*{ Preheating, Thermalization  and Supergravity}
\toctitle{Preheating, Thermalization  and Supergravity
\protect\newline  Preheating, Thermalization  and Supergravity}
%
%
\titlerunning{ Preheating, Thermalization  and Supergravity}
%
\author{Lev Kofman}
\authorrunning{Lev Kofman}
%
%
\institute{CITA, University of Toronto, M5S 3H8, Canada}

\maketitle              

\begin{abstract}
\index{abstract} This constribution collects 
 new recent  results  on preheating after inflation. 
 We discuss tachyonic preheating in the SUSY motivated
 hybrid inflation; development of  equilibrium after preheating;
 theory of fermionic preheating
and the problem of gravitino overproduction from preheating.
\end{abstract}

\section{Introduction}

\begin{figure}[b]
\centering \leavevmode\epsfysize=5.5cm \epsfbox{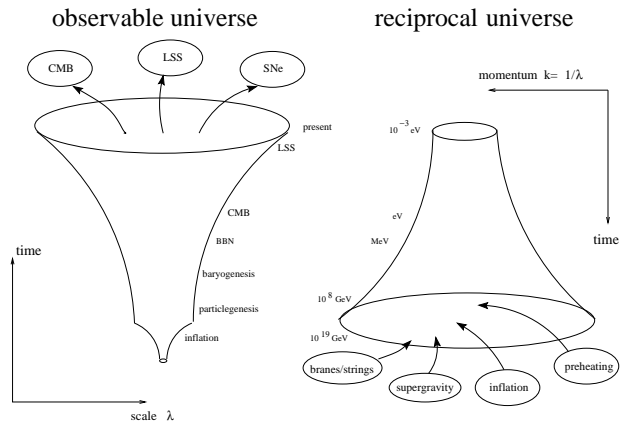}\\
\caption[fig1]{\label{fig1} 
Sketch of an expanding  universe where the wavelengths $\lambda(t)$ are 
 proportional to a  scale factor $a(t)$
 and  ``reciprocal'' universe where the momenta $k(t) \sim 1/\lambda(t) $
 are inversly proportional to  $a(t)$.}
\end{figure}

The best-fit universe is  the flat FRW   universe with
 scale-free  scalar metric  fluctuations.
The model is in agreement with  the observations  of the cosmic
microwave background  radiation (CMB) anisotropy, the large scale structure
(LSS) of the universe and the tests of the global geometry such as 
 the high redshift supernovae (SNe). The theoretical scenario of the expanding
 universe
includes  a very early stage  of inflation, subsequent stage of particle
creation during preheating and the setting of the thermal
 equilibrium.
The resulting hot expanding  universe passes through a sequence of crucial 
stages.
As the universe expands, it  opens up  for
 the observational tests,
beginning ``upward'' from the big bang nucleosynthesis (BBN),
see the left panel of Figure \ref{fig1}.

Although inflation is an essential part of the cosmological paradigm,
from the particle physics perspective it is not easy to construct a
 satisfactory model of the very early universe including
inflation. The right panel of  Figure \ref{fig1} shows a theoretical
``reciprocal'' universe where  the momenta $k \sim 1/\lambda$ are increasing
backwards in time.
The reciprocal  universe at the  earliest times at  high momenta 
 opens up 
for theoretical possibilities of the early universe physics,
including recent developments in brane cosmology, string motivated
cosmology, supergravity in cosmology etc. 

Fundamental M-theory  should encompass both supergravity
and string theory.
At present the low-energy phenomenology is described by the $N=1$ $d=4$
supergravity. Some preferable
choices of the K\"{a}hler  potentials, superpotentials and Yang--Mills
couplings hopefully will be selected at the level of the  fundamental theory.
Until the fundamental theory of all interactions is well understood,
one may try to address the issues of the eary universe  cosmology
in the context of the most general phenomenological $N=1$
supergravity--Yang--Mills--matter theory. This, in fact,
was the case during the last almost 20 years. 
A rather lengthy $N=1$ phenomenological supergravity Lagrangian 
begins with the terms
\begin{eqnarray}
e^{-1}{\cal L}&=&-{1 \over 2}M_P^2 R- 
\hat\partial_{\mu}  \Phi^i\hat\partial^{\mu}  \Phi_i+
 e^K\left({\cal D}^iW {\cal D}_iW  -3 {{WW^*} \over M_P^2} \right)\nonumber\\
&-&\bar \chi_j  \not\!\! {\cal D} \chi^i- \bar \chi^i
 \not\!\! {\cal D} \chi_j -\left( e^{K/2} {\cal D}^i{\cal D}^j  W\bar \chi_i
\chi _j + h.c. \right)\nonumber\\
&-& {1 \over 2}\bar \psi_\mu R^\mu +\left(
{1 \over 2}e^{K/2} W\bar \psi_{\mu R} \gamma ^{\mu \nu }\psi
_{\nu R}+ \bar \psi_{\mu L}{ \not\!\hat \partial } \Phi^i
 \gamma^\mu \chi_i+\bar  \psi_R\chi_i  e^{K/2} {\cal D}^iW+ h.c.
\right)\nonumber\\
&+&\,... \label{lag}
\end{eqnarray}
A particular  choice of the form of the Lagrangian  is motivated and 
notations are given in \cite{KKLV2}. In Eq. (\ref{lag})  $K$ is the 
 K\"{a}hler potential, $\Phi^i$ is the complex conjugate of $\Phi_i$. 

According to  the inflationary scenario,
 the Universe initially expands quasi-exponentially
in a vacuum-like state without entropy or particles.
At the stage of inflation, all energy is contained 
 in a classical slowly moving  fields $\Phi$ in the inflaton sector.
The last term of the 1st
line of (\ref{lag}) is the scalar potential $V(\Phi_i)$.
The equations of motion
 based on the first line
should describe inflation, which is a challenging problem by itself.
The Lagrangian  (\ref{lag})
contains also  other fields
which give subdominant contributions to gravity.
The Friedmann equation for the scale factor $a(t)$ and
the Klein-Gordon equation for $\Phi(t)$
determine the evolution of the background fields.
In  the chaotic inflation models, soon after the end of inflation,
an almost homogeneous inflaton field $\Phi(t)$   coherently
oscillates with a very large  amplitude of the order of the Planck mass
around the minimum of its potential. This scalar field can be considered as
a coherent superposition of inflatons  with zero momenta.
 The amplitude of  oscillations  gradually
decreases not only because of the expansion of the
universe, but also because energy is transferred to particles
created by the oscillating field.
At this stage
we shall recall the rest of the fundamental Lagrangian
  which
includes all the fields interacting with inflaton.
These interactions
  lead  to the  creation of many ultra-relativistic
particles from the inflaton.
Gradually, the inflaton field decays and transfers
all of its energy    to the created
particles.
In this scenario
 all the matter constituting the universe
is created from this process of reheating.
If the creation of particles is sufficiently slow,
   the  particles would   simultaneously
 interact with each other and come to a state of thermal equilibrium
at  the reheating temperature $T_R$.
This gradual reheating can be treated with the perturbative theory of
particle creation and thermalization.
However, typically particle production from coherently oscillating
inflatons occurs not in the perturbative regime but in the non-perturbative
regime of preheating \cite{KLS}. 

In this contribution I will discuss  several problems  of
preheating after inflation, some of them  related to the supergravity.
This discussion is based on new results derived in recent papers
 \cite{KKLV1,KKLV2,montr,FK,hybr,GK}.

\section{Tachyonic Preheating}

Another popular class of inflationary models --
hybrid inflation -- involve multiple scalar fields $\Phi_i$ in the inflaton 
sector.
Previous studies of preheating in hybrid models
were concentrated on particle creation by parametric resonance that may
occur when homogeneous background fields oscillate around the minimum of
the potential. Such parametric resonance may or may not be strong
depending on the coupling parameters. However, 
 we  recently found \cite{hybr} that there is strong preheating in
hybrid inflation, but its character is quite different from
preheating based on parametric resonance.
It turns out that
typically there is  tachyonic instability that
appears in a broad class of hybrid inflation models. The backreaction of
rapidly generated fluctuations does not allow homogeneous background
oscillations to occur because all energy of the oscillating field is
transferred to the energy of long-wavelength scalar field fluctuations
within a single oscillation.  However, this does not
preclude the subsequent decay of the Higgs and inflaton 
inhomogeneities  into
other particles, and thus reheating without parametric resonance.

Consider  the simple
potential for the two-field hybrid inflation is
\begin{equation} \label{hyb_eqn}
V(\phi, \sigma) = {\lambda \over 4} (\sigma^2 - v^2)^2 + {g^2 \over
2} \phi^2 \sigma^2  \, ,
\end{equation}
where we used notations  $\Phi_1=\phi$,  $\Phi_2=\sigma$.
Inflation in this model occurs while the homogeneous $\Phi_1$ field
slow rolls from large $\phi$ towards the bifurcation point at
$\phi = {\sqrt{\lambda} \over g} v $ (due to the slight lift of
the potential in $\phi$ direction).  Once $\phi(t)$ crosses the
bifurcation point, the curvature of the $\sigma$ field,
$m^2_{\sigma} \equiv \partial^2 V/\partial \sigma^2$, becomes
negative. This negative curvature results in exponential growth of
$\sigma$ fluctuations. Inflation then ends abruptly in a
``waterfall'' manner. 

One reason to be interested in hybrid inflation is that it can be
implemented in supersymmetric theories. In particular, for
illustration we will consider preheating in the 
 supersymmetric F-term inflation as an example
of a hybrid model. 

The simplest F-term hybrid inflation model (without undesirable domaine walls)
is based on a superpotential
with three  left-chiral superfields $\Phi_i=(\Phi_1, \Phi_2, \Phi_3)$
in the Lagrangian (\ref{lag})
\begin{equation}\label{super}
W = {\sqrt{\lambda}\over2}\Phi_1\left(4\Phi_2 \Phi_3 - v^2\right) \, .
\end{equation}
In this case, the spontaneous breaking of the local (global) $U(1)$ symmetry
between the $\Phi_2$ and $\Phi_3$ fields will lead to gauge (global)
string formation. 

In global SUSY, using the same notation for superfields and their
complex scalar components, this superpotential contributes 
\begin{equation} \label{pot_fterm}
V_{\rm F} = {\lambda\over4} |4\Phi_2\Phi_3 - v^2|^2 +
4\lambda |\Phi_1|^2 \left(|\Phi_3|^2 + |\Phi_2|^2\right) \, .
\end{equation}
to the scalar potential. In general, $\Phi_3$ and $\Phi_3$ could
be (oppositely) charged under a local $U(1)$ symmetry, in which case we
should include a D-term, $V_{\rm D}$,
which we negelect here. 
 
In this model, inflation occurs when chaotic initial conditions lead to
$\langle |\Phi_1 | \rangle \gg v$. When this happens, the  fields $\Phi_2$ and
$\Phi_3$ 
acquire large effective masses and roll to their local minimum at $\langle
\Phi_2\rangle = \langle\Phi_3\rangle =0$.  In this limit, the
potential~(\ref{pot_fterm}) becomes $V \approx {\lambda v^4 \over 4}$,
which gives rise to a non-vanishing effective cosmological constant.
However, this is a false vacuum state; the true vacuum corresponds to
$\langle\Phi_2\Phi_3\rangle = {v^2 \over 4}$ and $\langle \Phi_1
\rangle = 0$.  The slow-roll potential drives the evolution of
the inflaton towards its true VEV.  When its
magnitude reaches the value $\langle|\Phi_{\rm c}|\rangle = {v\over2}$
spontaneous symmetry breaking occurs.

For further discussion of symmetry breaking in this model, let us
rewrite~(\ref{pot_fterm}) in terms of
polar fields: $\Phi_3=| \Phi_3 | e^{i \theta}$,
${\Phi_2}=|\Phi_2  | e^{i {\bar \theta}}$.  The potential becomes
\begin{equation} \label{potff}
V_{\rm F} = {\lambda \over 4}\left( 16|\Phi_2|^2 |\Phi_3|^2 -
8v^2|\Phi_2||\Phi_3| \cos(\theta+\bar\theta) + v^4\right) +
4\lambda |\Phi_2|^2 \left( |\Phi_2|^2 + |\Phi_3|^2\right) \, .
\end{equation}
At the stage of symmetry breaking, when $\langle | \Phi_2 \Phi_3 | \rangle$
begins to move away from zero, the absolute phase
${\rm Arg}(\Phi_2 \Phi_3 ) = \theta+\bar\theta$ acquires a mass and is
forced to zero.  Note, however, that the potential is independent of the
relative phase, $\theta-\bar\theta$, reflecting the $U(1)$ symmetry.
Thus, in a quasi-homogeneous patch, the $U(1)$
symmetry allows us to choose the relative phase of the $\Phi_2$, $\Phi_3$
 fields
to be zero without any loss of generality.  This choice, combined with the
vanishing of the absolute phase, is equivalent to choosing the two complex
$\Phi_2$, $\Phi_3$  fields to be real. In order to leave
 canonical kinetic terms,
we define $\sigma_\pm \equiv |\Phi_3| \pm |\Phi_2|$.  Furthermore,
as inflation has left the inflaton homogeneous across all the patches, we
may choose it to be real: $\phi \equiv \sqrt2|\Phi_1|$.

In terms of these three real fields, the potential now becomes
\begin{equation}\label{potf}
V_{\rm F} = {\lambda \over 4} \left(\sigma_+^2 - \sigma_-^2 -
v^2\right)^2 + \lambda \phi^2 \left(\sigma_+^2 + \sigma_-^2\right) \ .
\end{equation}
In the symmetric phase, when $\sigma_\pm = 0$, the $\sigma$ fields
have an effective mass-squared: $m_\pm^2(\phi) = \lambda
\left( 2 \phi^2 \mp v^2 \right)$. 
We can now see that spontaneous symmetry breaking
 occurs in this model exactly as in the
two field model (\ref{hyb_eqn}).  For
$\phi < \phi_c = {v \over \sqrt{2}}$, the $\sigma_+$ field has a
tachyonic mass that triggers symmetry breaking and the end of inflation.
On the other hand, the $\sigma_-$ field
has always a large and positive effective mass-squared, pinning it
to zero.  Thus, during inflation and at the {\em initial} stages of symmetry
breaking, this model behaves just like the standard 
two field hybrid model discussed in the last section.  We have only to apply
the constraint $g^2 = 2\lambda$ and identify the Higgs field
with $\sigma_+$.
The equations for the homogeneous background components
$\phi(t)$ and
$\sigma_\pm(t)$ admits simple solution
\begin{eqnarray}\label{combination}
 \phi(t)+{ 1 \over \sqrt{2}}\sigma_+(t)=\phi_c \ ,  \hspace{2cm}
 \sigma_-(t)=0 \ .
\end{eqnarray}

To study preheating in the F-term inflation, we have to
 analyse evolution of the  vacuum fluctuations.  Consider  vacuum
fluctuations   in the inflaton sector $\Phi_i$ of the theory (\ref{potff}). 
Usual description in terms of  a homogeneous background plus small
fluctuations gives us equations for fluctuations around the background
solution  (\ref{combination}). We  define the  
variances of fields:
\begin{equation}
\left< \,  \vert \sigma_\pm - \left< \sigma_\pm \right> \vert^2 \,
\right>_{\rm ren} = \int{d^3k \over {(2 \pi)^3}} \, \left[ \,
\vert \delta\sigma_{k\pm}(t) \vert^2 - \vert \delta\sigma_{k\pm}(0) \vert^2
\, \right] \equiv \int{{{dk} \over k}} \, {\mathcal P}_{\pm}(k,t)
\end{equation}
and similar for $\phi$ field. 
Here $ {\mathcal P}_{\pm,\phi}(k,t)$ are the spectra of the fluctuations.

\begin{figure}[t]
\centering \leavevmode \epsfxsize=7.5cm    
\epsfbox{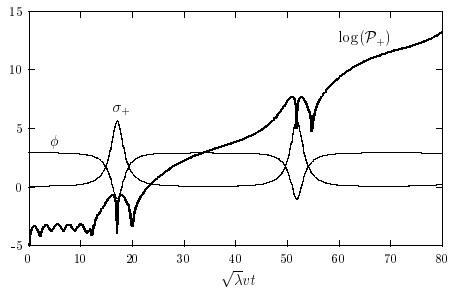}\\
\caption{Evolution of the background fields
 $\sigma_+(t)$ and $\phi(t)$ after symmetry
breaking and the $\log$ of ${\mathcal P}_+(k)$ for the mode with the momentum
$k = 0.2 \sqrt{\lambda} v$.  
} \label{bkgd}
\end{figure}

\begin{figure}[b]
\centering \leavevmode \epsfxsize=7.5cm
\epsfbox{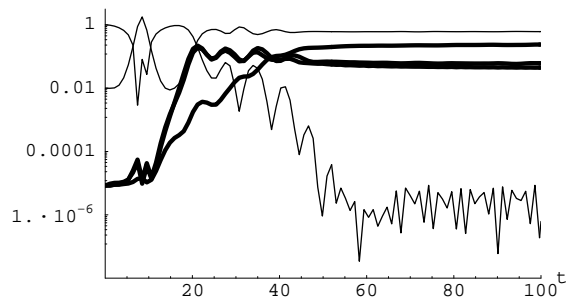}\\
\caption{Means and variances in units of $\phi_c$.
 The squared means $\langle\phi\rangle^2$
and $\langle \sigma_+ \rangle^2$ are ordinary solid lines while the field
variances  are thick lines. The mean of
$\phi $ starts at $\phi_c$, oscillates once, and then decays.
 The mean of $\sigma_+$ grows in antiphase
to $\phi$ and freeses at $\phi_c$.} \label{potmeansvars}
\end{figure}

 Numerical solutions of the equations  for
the background fields and 
  for the time
evolution of the  fluctuations of $\sigma_+$
for the mode $k$ where their spectrum is maximum
 are plotted as
the bold line at the Figure  \ref{bkgd}. 
Notice  an enormous exponential
growth of the fluctuations within a single background
oscillation. Indeed, the amplitude  ${\mathcal P}_+(k)$ increases by
factor $10^{10}$! 
 There are two factors which contributes to such a
strong  instability of fluctuations in the model. First,
oscillating background fields are crossing the region with significant
negative curvature of the effective potential, which
results in  tachyonic instability.
  Second, this region
turn out to be a turning point for the background oscillations, where
the fields spend significant portion of the oscillation.  As a result,
tachyonic instability is lasting long enough to make  the
backreaction of the fluctuations to be significant already within
single background oscillation.  The regime of background oscillations
 will even not be settled.
Therefore practically from the beginning we have to
use the lattice simulations to study  nonlinear
dynamics of the fields.
The results of full lattice simulations in the model 
are plotted in Figure \ref{potmeansvars}.
The simulations showed that the homogeneous fields
${\phi}$ and ${\sigma}_+$ initially followed the classical trajectory
(\ref{combination}) but, within one oscillation of the inflaton field,
fluctuations grew too large to speak meaningfully of the fields as
homogeneous oscillators. These fluctuations grew in such a way
that ${\sigma_+}={{\sigma_-}}^*$ almost exactly
throughout the simulation. In other words
Re $\delta{\sigma}_+$ and Im $ \delta {\sigma}_-$ were
excited while Im $ \delta {\sigma}_+$ and
Re $\delta {\sigma}_-$ were not. Because of this we only plot
the fields ${\phi}$ and ${\sigma_+}$.

In \cite{hybr} we develop a general theory of tachyonic
 preheating, which occurs due to
tachyonic instability in the theories with spontaneous symmetry breaking.
Our approach combines analytical estimates with lattice simulations
taking into account all backreaction effects. The process of spontaneous
symmetry breaking involves transfer of the potential energy into the
energy of fluctuations produced due to the tachyonic instability. We show
that this process is extremely efficient and requires just a single
oscillation of the scalar field falling from the top of the effective
potential.
We considered  preheating in the hybrid
inflation scenario, including SUSY motivated F-term and D-term
inflationary models.

\section{Development of Equilibrium}

The character of preheating may vary from model
to model, e.g. parametric excitation in chaotic inflation
\cite{KLS} and tachyonic  preheating  in hybrid
inflation \cite{hybr}, but its distinct feature remains the same:
rapid amplification of one or more bosonic fields to exponentially
large occupation numbers. This amplification is eventually shut
down by backreaction of the produced fluctuations. The end result
of the process is a turbulent medium of coupled, inhomogeneous,
classical waves far from equilibrium. 
Despite the development of our understanding of preheating after
inflation, the transition from this stage to a  hot Friedmann
universe in thermal equilibrium has remained relatively poorly
understood.
The details of this thermalization stage
depend on the constituents of the fundamental Lagrangian 
(\ref{lag})  and their
couplings, so at first glance it would seem that a description of
this process would have to be strongly model-dependent.
Recently we performed  a fully nonlinear study of the development of
equilibrium after preheating \cite{FK}.
We have
performed lattice simulations of the evolution of interacting
scalar fields during and after preheating for a variety of
inflationary models. 
 We have
found, however, that many features of this stage seem to hold
generically across a wide spectrum of models. 
Indeed, at the end of preheating and beginning of the turbulent
stage $t_*$, the fields are out of equilibrium. We
have examined many models and found that at $t_*$ there is not
much trace of the linear stage of preheating and conditions at
$t_*$ are not qualitatively sensitive to the details of inflation.
We therefore expect that this second, highly nonlinear, turbulent
stage of preheating may exhibit some universal, model-independent
features. Although a realistic model would include one or more Higgs-Yang-Mills
sectors, we treat the simpler case of interacting scalars.

 We have numerically
investigated the processes of preheating and thermalization in a
variety of models and determined a set of rules that seem to hold
generically. These rules can be formulated as follows
(in this section we use notations $\phi=\Phi_1$ for the inflaton field
and $\chi$, $\sigma$ for other scalars $\Phi_i$)

\bigskip
\noindent {\it 1.  In many, if not all viable models of inflation there
exists a mechanism for exponentially amplifying fluctuations of at
least one field $\chi$. These mechanisms tend to excite
long-wavelength excitations, giving rise to a highly infrared
spectrum.}

The mechanism of parametric resonance in single-field models of
inflation has been studied for a number of years. 
This effect is quite robust. Adding
additional fields (e.g.  $\sigma$ fields) or self-couplings
(e.g. $\chi^4$) has little or no effect on the resonant period.
Moreover, in many hybrid models a similar effect occurs due to
tachyonic  instability. The qualitative features of the fields
arising from these processes seem to be largely independent of the
details of inflation or the mechanisms used to produce the fields.

\bigskip
\noindent {\it 2.  Exciting one field $\chi$ is sufficient to
rapidly drag all other  light fields with which $\chi$
interacts into a similarly excited state.}

We have seen this effect when multiple fields are coupled directly to
$\chi$ and when chains of fields are coupled indirectly to $\chi$. All
it takes is one field being excited to rapidly amplify an entire
sector of interacting fields. These second generation amplified fields
will inherit the basic features of the $\chi$ field, i.e. they will
have spectra with more energy in the infrared than would be expected
for a thermal distribution.

\bigskip
\noindent {\it 3. The excited fields will be grouped into subsets with
identical characteristics (spectra, occupation numbers, effective
temperatures) depending on the coupling strengths.}

We have seen this effect in a variety of models. For example in
the models (\ref{nfldlambda}) which we are going to consider
  the $\chi$ and
$\sigma$ fields formed such a group. In general, fields that are
interacting in a group such as this will thermalize much more
quickly than other fields, presumably because they have more
potential to interact and scatter particles into high momentum
states.

\bigskip
\noindent {\it 4. Once the fields are amplified, they will approach
thermal equilibrium by scattering energy into higher momentum modes.}

This process of thermalization involves a slow redistribution of
the particle occupation number as low momentum particles are
scattered and combined into higher momentum modes. The result of
this scattering is to decrease the tilt of the infrared portion of
the spectrum and increase the ultraviolet cutoff of the spectrum.
Within each field group the evolution proceeds identically for all
fields, but different groups can thermalize at very different
rates.

Here we will illustrate these results with
a simple  chaotic inflation model with a
quartic inflaton potential. The inflaton $\phi$ has a four-legs
coupling to another scalar field $\chi$, which in turn can couple
to two other scalars $\sigma_1$ and   $\sigma_2$. The potential for this
model is
\begin{equation}\label{nfldlambda}
V = {1 \over 4} \lambda \phi^4 + {1 \over 2} g^2 \phi^2 \chi^2 +
{1 \over 2} h_1^2 \chi^2 \sigma_1^2+{1 \over 2} h_2^2 \chi^2 \sigma_2^2
\end{equation}
Preheating in this theory in the absence of the $\sigma_i$ fields
is well studied.  For nonsmall $g^2 \over \lambda$ the field
$\chi$ will experience parametric amplification, rapidly rising to
exponentially large occupation numbers. In the absence of the
$\chi$ field (or for sufficiently small $g$) $\phi$ will be
resonantly amplified through its own self-interaction, but this
self-amplification is much less efficient than the two-field
interaction. The results shown here are for $\lambda = 9 \times
10^{-14}$ (for CMB anisotropy normalization) and $g^2 = 200 \lambda$. When
we add a third field we use $h_1^2 = 100 g^2$ and when we add a
fourth field we use $h_2^2 = 200 g^2$.

One of the most interesting variable to calculate is the
(comoving) number density of particles of the fields $n(t)$
and their  occupation number $n_k$.
The evolution
of the total number density of all particles $n_{tot}$ is an indication of the
physical processes taking place. In the weak interaction limit the
scattering of classical waves via the interaction  ${1 \over
2} g^2 \phi^2 \chi^2$ can be treated using a perturbation
expansion with respect to $g^2$. 
The leading four-legs diagrams
for this interaction corresponds to a two-particle collision
$(\phi \chi \rightarrow \phi \chi)$, which conserves $n_{tot}$.
The regime where such interactions dominate corresponds to ``weak
turbulence'' in the terminology of the theory of wave turbulence.
 If we see $n_{tot}$ conserved it will be an
indication that these two-particle collisions constitute the
dominant interaction. Conversely, violation of $n_{tot}(t)=const$
will indicate the presence of strong turbulence, i.e. the
importance of many-particle collisions. Such higher order
interactions may be significant despite the smallness of the
coupling parameter $g^2$ (and others) because of the large
occupation numbers $n_k$. Later, when these occupation numbers are
reduced by rescattering, the two-particle collision should become
dominant and $n_{tot}$ should be conserved.
For a bosonic field in thermal equilibrium with a temperature $T$
and a chemical potential $\mu$ the spectrum of occupation numbers
in the limit of classical waves 
is given by
\begin{equation}\label{wave}
n_k \approx {T \over {\omega_k -\mu}} \, .
\end{equation}

\begin{figure}[t]
\centering \leavevmode \epsfxsize=8.0cm
\epsfbox{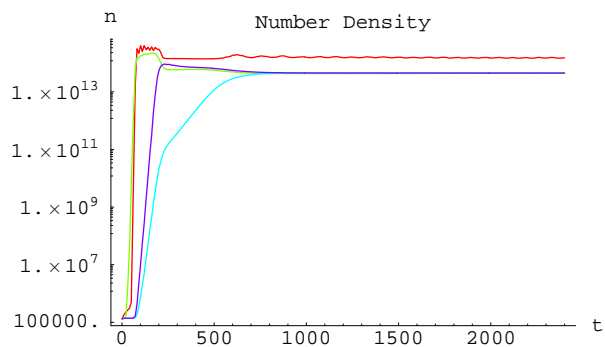}\\
\caption{ Time  evolution of number density of particles in the model
(\ref{nfldlambda}). The curves represent $n_{\phi}$,  $n_{\chi}$,
 $n_{\sigma_1}$, $n_{\sigma_2}$ from top to bottom. Unit of
(conformal) time is  $a \cdot 10^{-36}$ sec.} \label{n4fldl}
\end{figure}

\begin{figure}[b]
\centering \leavevmode \epsfxsize=8.0cm
\epsfbox{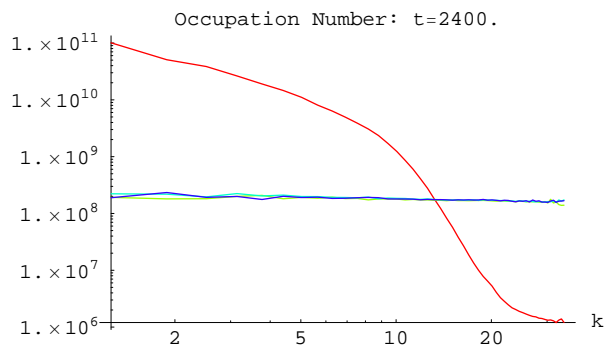}
\caption{Occupation number versus momentum for the model
(\ref{nfldlambda}).  The curves represent spectra of particles for the fields
$\phi, \chi, \sigma_1$ and $\sigma_2$ from top to bottom
All the fields other than the inflaton have nearly identical spectra}
\label{nk4fldl}
\end{figure}

Figure \ref{n4fldl} shows an exponential increase
of  $n(t)$ during preheating, followed by a gradual decrease
 that asymptotically slows down.
This exponential increase is a consequence of
explosive particle production due to parametric resonance.
After preheating the
fields enter a turbulent regime.  In our simulations we see
$n(t)$ decreasing during this stage.
 This decrease is a consequence of the
many-particle interactions beyond the four legs rescattering.

Figure \ref{nk4fldl} 
illustrates the late-time spectrum of particles.
Initially spectra are concentrated at the infrared section,
which  is gradually flattening as it
approaches a thermal distribution (\ref{wave})
 (i.e. a slope of $-1$ to $0$ depending on the chemical
potential  and the mass).
The spectra of three fields are essentially identical, which leads to
\begin{equation}\label{four}
n_{\chi} \approx  n_{\sigma1}  \approx  n_{\sigma2}  < n_{\phi} .
\end{equation}
 Another important point is that
the interaction of $\chi$
and $\sigma_i$ does not affect the preheating of $\chi_i$, but does
drag $\sigma_i$ exponentially quickly into an excited state.
 The fields $\sigma_i$ are exponentially amplified not by
parametric resonance, but by their stimulated interactions with the
amplified $\chi$ field. Unlike amplification by preheating, this
direct decay nearly conserves particle number, with the result
that $n_\chi$ decreases as $\sigma_i$ grow.

\begin{figure}[b]
\centering \leavevmode \epsfxsize=8.0cm
\epsfbox{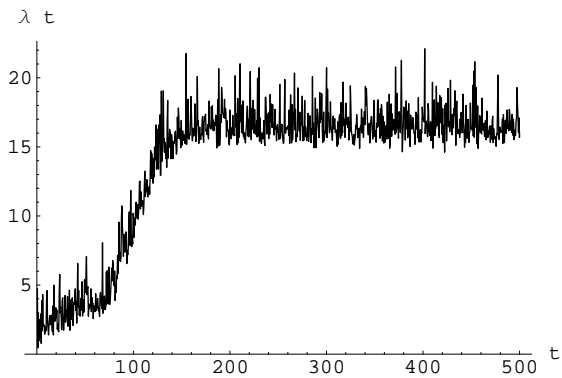}
\caption{The Lyapunov exponent
$\lambda$ for the fields $\phi$ and $\chi$ using the normalized
distance function $\Delta$.
}\label{lyapunovdelta}
\end{figure}

Interacting waves of scalar fields constitute a dynamical system.
 Dynamical chaos is one of the features of wave
turbulence. In \cite{FK}  we address the question  how and
when the onset of chaos takes place after preheating.
To investigate the
onset of chaos  we have to follow the time evolution
of two initially nearby points in the phase space.
Consider the theory with the potential
(\ref{nfldlambda}) with two fields $\phi$ and $\chi$ only
(which we collectively denote as $f$).
Consider two configurations of a scalar field $f$ and
$f'$ that are identical except for a small difference of the
fields at a set of points $x_A$.
Chaos can be defined as the tendency of such
nearby configurations in phase space to diverge exponentially over
time.
This divergence is parametrized by the Lyapunov exponent for
the system, defined as
\begin{equation}
\lambda \equiv {1 \over t} log {\Delta(t)\over \Delta_0}
\end{equation}
where $\Delta$ is a distance between two configurations 
and $\Delta_0$ is
the initial distance at time $0$.
In the context of preheating is is convenient to
define $\Delta$ as
\begin{equation}\label{ratio}
\Delta(t) \equiv \sum_A \left({{f'_A-f_A}\over {f'_A+f_A
}}\right)^2 +\left({{\dot f'_A-\dot f_A}\over {\dot f'_A+\dot f_A
}}\right)^2
\end{equation}
that is well regularized even while the field $\chi$ is being
amplified exponentially during preheating.
Figure~\ref{lyapunovdelta} shows the
Lyapunov exponent $\lambda$.
 We see the onset of
chaos only at the end of preheating. The plot for the $\phi$ field
is nearly identical.  The Lyapunov exponents for the fields were
$\lambda_\phi \approx \lambda_\chi  \approx 0.2$ (in the units
of time adopted in the simulation). This corresponds to a very
fast onset of chaos around the moment $t_*$ where
the strong tubulence begins.

The highlights of our study for early universe phenomenology are
the following. The mechanism of preheating after inflation is
rather robust and works for  many different systems of
interacting scalars. There is a stage of turbulent classical waves
where the initial conditions for preheating are erased. Initially,
before all the fields have settled into equilibrium with a uniform
temperature, the reheating temperature may be different in
different subgroups of fields. The nature of these groupings is
determined by the coupling strengths.

\section{Parametric Excitations  of Fermions}

The interaction between bosons and inflaton(s)
may results in the copious production -- preheating --
 of the bose particles,
either due to the parametric resonance as in the
chaotic inflation or due to the tachyonic instability
as in the hybrid inflation. Preheating of bosons has been  studied 
in details with  analytic approximations
 as well as with the lattice simulations. 
 Consider second line of Eq.~(\ref{lag})
and ignore mixing between fermions  $\chi$ and gravitino 
$\psi_{\mu}$ (which corresponds to the
rigid SUSY limit). There is  interaction between 
inflaton and fermions. Let us consider a simple 
  Yukawa interaction  $h\bar \psi \phi \psi$, which should
lead to the production of fermions from inflaton oscillations
(in this section we use notation $\phi$ for the inflaton field).
For fermions, the Pauli exclusion principle prohibits the 
occupation number from exceeding $1$.  For this reason,
it has  been silently assumed that  fermions are created  in the three-legs
perturbative process  $\phi \to  \bar  \chi  \chi$
 where individual inflatons  decay independently into 
 pairs of  $\chi$-particles.
Following \cite{GK} let us, however, consider the Dirac equation for a
massless quantum Fermi field $\chi(t, \vec x)$ intracting
with the background inflaton 
\begin{equation}
\left[  \gamma^{\mu} \nabla_{\mu} +h \phi(t)\right] \chi=0 \ ,
\label{5}
\end{equation}
where $ \nabla_{\mu}$ is the derivative with the spin connection.
Here, similar to the bosonic case, the  inflatons producing fermions  also 
 act not as individual particles  but as a coherently oscillating
field $\phi(t)$. 
Indeed, the equation  for the eigenfunctions of the 
quantum fluctuations in this theory 
 can be reduced to a second-order equation for an
auxillary field $X(t, \vec x)$,
so that
 $\Psi= \left[  \gamma^{\mu} \nabla_{\mu} +h \varphi\right]
 X$. The eigenmodes of the auxillary field have the form
$ X_k(t)e^{ +i{{\bf k}} \cdot {{\bf x}}}R_r$, 
where the $R_r$ are  eigenvectors
of the Dirac matrix $\gamma^0$ with eigenvalue $+1$
(we are using the representation of gamma matrices where
 $\gamma_0=diag(i, i, -i, -i)$).
The temporal part of the eigenmode
obeys an oscillator-like equation with a complex frequency
which depends periodically on time
\begin{equation}
\ddot  X_k  +  {\left(k^2  +\Omega^2-i \dot \Omega
 \right)} X_k  = 0 \ ,
\label{100}
\end{equation}
where $\Omega= \sqrt{q}  f$.
The background oscillations enter in the form
 $f(t)={\phi(t) \over \phi_0}$, where $\phi_0$ is their  amplitude.
Without imaginary part, this equation will the the equation for the
fluctuations of the bose field, which results in its parametric resonance.
The imaginary part of the frequency in  Eq.~(\ref{100}) guarantees
the Pauli blocking for the occupation number $n_k$.
The results for  $n_k$ can be formulated as
follows \cite{GK}.
Even though the Yukawa interaction  contains a small
factor $h$, one cannot use the perturbation
expansion in ${h}$.
The strength of the effect is ultimately determined
by the dimensionless parameter $q$. For instance, for chaotic inflation with
$\lambda \phi^4$ potential, $q={h^2 \over \lambda}$
is typically very large. For instance, in the simple supergravity theory
with $W=\sqrt{\lambda}\phi^3$ and single chiral multiplet,
$h=\sqrt{2\lambda}$ and $q=2$.
For the  quadratic inflaton potential
$m^2 \phi^2$,  the parameter
$q={{h \phi_0} \over M_p}$, where $\phi_0$ is the amplitude
of the inflaton oscillations, which initially is $\sim 0.1 M_P$.
 \begin{figure}[t]
   \centering \leavevmode \epsfxsize=8.0cm
   \epsfbox{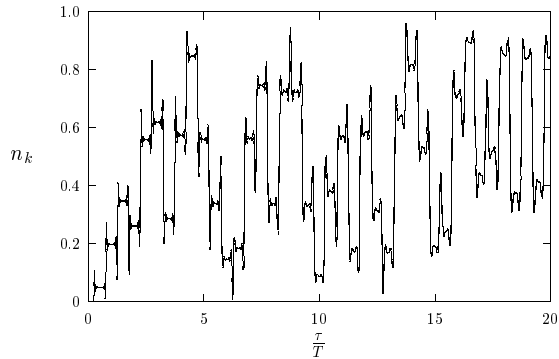}\\
   \caption[fig1u]{\label{stoch} {\em 
The occupation number $n_k$ of fermions in $m_\phi^2 \phi^2$-inflation.
  The initial resonance parameter
is $q_0=10^6$ and the mode is $\kappa^2_0 \approx \sqrt{q}m^2$,
 $T={{2\pi} \over m}$.}}
\end{figure}
\begin{figure}[b]
   \centering \leavevmode \epsfxsize=8.0cm    
   \epsfbox{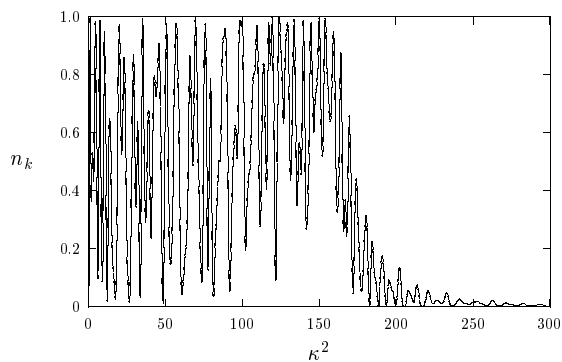}\\
 \caption[fig1u]{\label{sphere} {\em 
The comoving occupation number of fermions in $m_\phi^2 \phi^2$-inflation 
after $50$ inflaton oscillations for initial resonance parameter $q_o = 10^3$.
Expansion of the universe destroys the details of the resonance band and
leads to a fermi-sphere of width $q^{1/4}m$.}}
\end{figure}
The modes
get fully excited with occupation numbers $n_k \simeq 1$ 
within
 tens of oscillations of the field $\phi$, and the width of the
parametric excitation of fermions in momentum space is about
$q^{1/4} m_{eff}$, where  $m_{eff}$ is the frequency of the background 
  inflaton oscillations.
It turns out that the analytic  theory of fermionic preheating can be 
advanced similar to that for the bosonic preheating, see  \cite{GK}
for details. For illustration in 
Figure \ref{stoch} we show the  time evolution of the occupation number
of fermions in $m^2\phi^2$ inflation, where excitations  of fermions is
 a stochastic process. Figure  \ref{sphere} shows the spectrum 
of fermionic excitations, which stochastically filling a (Fermi) sphere
in the momentum space \cite{GK}.

\section{Gravitino from Preheating}

Let us now consider the third line in the supergravity Lagrangian (\ref{lag}),
which describes the gravitino field $\psi_{\mu}$.
Assume that the gravitino mass $m_{3/2} \sim 10^2-10^3$Gev. 
Such particles decay
 very late after  nucleosynthesis and lead to disastrous cosmological
 consequences unless the abundance of gravitinos 
(in units of the  number density of photons) is extremely suppressed, 
$n_{3/2}/n_{\gamma} < 10^{-15}$. 
 This puts constrains on the reheat temperature $T_R$ to avoid 
gravitino overproduction from thermal scattering of particles 
in the very early universe, $T_R < 10^8$ Gev.
However, gravitinos can be produced not only in the thermal bath
after preheating but even earlier during the inflaton 
oscillations, similarly to the production of fermions from
preheating \cite{KKLV1,GTR}.
Thus the investigation of the non-thermal gravitino production in
the early universe may serve as a useful tool  to
discriminate among various versions of inflations.

To study gravitino production from preheating, first one shall derive
its equation, which is a challenge by itself. The formalism for gravitinos
in  an expanding universe  and in the presence of complex scalar
fields with non-vanishing VEV's was  recently formulated in paper 
\cite{KKLV2}.
The  equation for the
gravitino  has on the left hand side  the kinetic part $R^{\mu} 
\equiv \epsilon^{\mu\nu\rho\sigma}\gamma_5\gamma_{\nu}
{\cal D}_{\rho}\psi_{\sigma}$, 
 and a rather lengthy right hand side
 (we  use the long
derivative ${\cal D}_{\mu}$ with the spin connection and Christoffel
symbols, for which ${\cal D} _\mu \gamma_\nu=0$).
Apart of varying
gravitino mass $m_{3/2}={|W| \over M_p^2}$, the right hand side
contains a chiral
 connection  and various mixing terms
like those in the 3rd  line of (\ref{lag}).
 For a self-consistent setting of the
problem, the gravitino equation should be supplemented by the
equations for the fermions  $\chi_i$ and gauginos $\lambda_{\alpha}$
 mixing with gravitino,
 as well as by the
equations determining the gravitational background and the
evolution of the scalar fields.
We formulated  cosmological extension of the  super-Higgs effect
\cite{KKLV2}. It turns out that supersymmetry is spontaneously broken
if the combination
\begin{equation}
H^2 + m_{3/2}^2  > 0
\end{equation}
is positive. In flat space-time usually supersymmetry breaking is associated
with the non-vanishing gravitino mass $m_{3/2}$. In an expanding universe
 the Hubble parameter $H$ plays an equally important role. For instance, the 
gravitino mass (i.e. superpotential $W$) can vanish, but 
supersymmetry  will be still spontaneously broken.
For further discussion of the nonthermal gravitino production, 
let us make some simplifications. We consider the supergravity
multiplet and a single chiral multiplet
 containing
a complex scalar field $\Phi$ with   a single chiral fermion $\chi$.
This is a simple non-trivial extension which allows us to study the
gravitino with a non-trivial FRW  cosmological metric
 supported by the scalar field.
 A nice feature of this model is that
   the chiral fermion $\chi$
can be gauged to zero so that the mixing between $\psi_{\mu}$ and
$\chi$ in (\ref{lag}) is absent. We also can choose the
non-vanishing VEV of the scalar field (inflaton) in the real direction.
First we will derive the equation for a spin 3/2 field in a
curved  background metric with non-vanishing VEVs for the scalar fields.
From (\ref{lag})  we can obtain  the  equation for the gravitino 
 We use a plane-wave
ansatz $\psi_{\mu} \sim e^{i {\bf k \cdot x}}$ for the
space-dependent part. Component $\psi_0$ is related to $\psi_i$ via
the constraint equation.
 Next,  $\psi_i$ can be decomposed into
  its  transverse part
$\psi^T_i$, and to the longitudinal part $\psi^L_i$ which is defined by the
 trace $\gamma^i\psi_i$.
 Two degrees of freedom of $\psi_{\mu}$ are
associated with the transverse part $\psi^T_i$, which correspond
to helicity $\pm3/2$ and two degrees of freedom are associated with
 $\gamma^i\psi_i$ (or $\psi_0$)
which correspond to helicity $\pm1/2$.
Equation for  the helicity $\pm 3/2$ states 
 for $\mu=i$ is 
\begin{equation}
\left(\gamma ^{\mu} 
\partial _\mu + \frac{ a'}{2a}\gamma^0 +
 m_{3/2}a \right)\psi_i^T=0 \,.
\label{trans}
\end{equation}
 The transformation
$\psi^T_i=a^{-1/2}\Psi^T_i$ reduces
 the equation for the transverse part to the
free Dirac equation with a time-varying mass term $m_{3/2}a$, c.f. eq.
 (\ref{5}).
In the previous section we explained 
 how to treat this type of equation.
The essential part  of $\Psi^T_i$ is given by the time-dependent
part of the eigenmode  of the transversal component $X_T(t)$,
which obeys second-order equation (c.f. (\ref{100})):
\begin{equation}
\ddot X_T +\left(k^2+(m_{3/2}a)^2-i\dot(m_{3/2}a)\right)X_T=0 \ .
 \label{dirac}
\end{equation}
Production of helicity 3/2 gravitino is essential at the fast rolling 
stage of inflaton evolution, when $W$ may be changed nonadiabatically.

The corresponding equation for gravitino with helicity 1/2 is more
complicated. It is convenient to use combination  $\theta=\gamma^i\psi_i$
which fully describes helicity $1/2$ states.
 Equation for combination 
The  equation for $\theta$ is 
\begin{equation}
 \left( \partial _t +\hat{B}-
i{\bf k  \cdot  \gamma} \gamma _0 \hat{A} \right)
 \theta=0\,,
 \label{trace1}
\end{equation}
where
$ \hat  B =-\frac{3\dot a}{2a}\hat{A}-{ {m a} \over 2}\gamma _0(1
+3\hat{A})$, and
\begin{equation}
  \hat A= {{ p-3m^2M^2_P}
 \over { \rho+ 3  m^2M^2_P }}\,
+\,\gamma_0{{2 \dot m M^2_P}
 \over { \rho+ 3  m^2 M^2_P}} =A_1+\gamma _0A_2\,.
\label{A2}
\end{equation}
Here $\rho$ and $p$
are the  energy-density and pressure of the background scalar field.
For the single scalar field 
$\vert A \vert^2 =1$
for an arbitrary superpotential $W$.
Thus  $ A$  can be represented as
 $ A =-\exp \left(2{ i}\int_{-\infty }^t dt \,\mu (\eta)\right)$, where
$ \mu ={\cal D}{\cal D}W + \Delta$,
the correction  $\Delta={\cal O}(M_P^{-1})$ is given in \cite{KKLV2}.
The time-dependent factor of the spinor $\theta$, which we denote as
$f_k(t )$, obeys a second-order differential equation.
  By the
substitution $f_k(t ) = E(t)X(t)$, with
$E=(-A^*)^{1/2} e^{-
 \int dt\, Re B},$
  the equation   for the function $f_k(t )$  is reduced to the
 oscillator-like equation for the time-dependent mode
function $X_L$ of the longitudinal component:
\begin{equation}
\ddot X_L+\left(k^2+\Omega_L^2-i\dot \Omega_L\right)X_L=0 \,
 \label{final}
\end{equation}
with
$a^{-1} \Omega_L=
\mu-\frac{3}{2} H\sin 2{\int \mu dt }- \frac{1}{2}
m \left(1 +3 \cos{2\int \mu dt }\right)$.
Equation (\ref{final})  describes
the creation of gravitinos from preheating.
Notable  the gravitino with
helicity 1/2  remains coupled to the
changing background even in the limit $M_P \to \infty$. In a
sense, the gravitino with helicity 1/2 remembers its goldstino
nature. Thus gravitino production in
this background in general is not suppressed by the gravitational
coupling. The main dynamical quantity which is responsible for the
gravitino production in this scenario will not be the small
changing gravitino mass $m_{3/2}$, but the mass of the chiral
multiplet $\mu$, which is much larger than $m_{3/2}$.

But do we really produce gravitinos from preheating, or we
describe the production of the chiral fermion?
To answer this question, we have to investigate more realistic and more
complicated problem with several chiral multiplets.
As the first step in this direction we derived equations for
gravitinos for an arbitrary number of chiral multiplets plus SYM sector
\cite{KKLV2}. For instance, generalization of the equation (\ref{trace1})
for the next simplest case of two chiral multiplets even in the limit
$M_P \to \infty$ is reduced to the complicated second-order spinor equation
\begin{equation}
\left( \partial_t +
{\hat B}-i{\bf k  \cdot  \gamma} \gamma_0 {\hat A} \right)\theta
+{\hat G}  \left[ \partial_t^2 +
(k^2 -i{\bf k  \cdot  \gamma}\gamma_0  \dot {\hat A}) \right] \theta =0\, ,
\label{trace2}
\end{equation}
the matrix ${\hat G}$ is constructed from the backgroud scalars.
We expect solutions of this  equations 
 will allow us to find the residual gravitino production.
Recent report \cite{LK} indicates that helicity 1/2 gravitinos 
are produced in the  model with two chiral multiplets.

%

\end{document}